\documentclass[journal]{IEEEtran}

\ifCLASSINFOpdf
   \usepackage[pdftex]{graphicx}
   \else
\fi

\usepackage{url}
\usepackage{graphicx}
\usepackage{amsmath}
\usepackage{pstricks-add}
\usepackage{pst-xkey}
\usepackage{pst-arrow}
\usepackage{mathtools}
\usepackage{mathptmx}
\usepackage{caption}
\usepackage{subcaption}
\usepackage{bbm}
\usepackage{tabstackengine}
\usepackage{mathtools}
\usepackage{amsmath}

\usepackage{amssymb}
\usepackage{bm}
\usepackage{float}

\usepackage{flushend}

\usepackage[utf8]{inputenc}

\begin{document}

\title{Reference Signal-Aided Channel Estimation in Spatial Media-Based Modulation Systems 
}

\author{Akif Kabac{\i}, Mehmet Ba{\c s}aran, \IEEEmembership{Member, IEEE,} and Hakan Ali {\c C}{\i}rpan, \IEEEmembership{Member, IEEE}

	
\thanks{Manuscript received XX XX, 2020; revised XX XX, 2020; accepted
	XX XX, 2020. Date of publication XX XX, XX; date of current version
	XX XX, 2020.}
\thanks{Akif Kabac{\i} and Hakan Ali {\c C}{\i}rpan are with Department of Electronics and Communication Engineering, Istanbul Technical University, Sariyer, Istanbul, 34467, Turkey (e-mail: kabaci@itu.edu.tr, cirpanh@itu.edu.tr)}%
\thanks{Mehmet. Ba{\c s}aran is with Information and Communications Research Group, Informatics Institute, Istanbul Technical University, Sariyer, Istanbul, 34467, Turkey (e-mail: mehmetbasaran@itu.edu.tr)}
\thanks{Digital Object Identifier }

}

\maketitle

\begin{abstract}
In this letter, we study the reference signal-aided channel estimation concept which is a crucial requirement to address the realistic performance of spatial media-based modulation (SMBM) systems where the radio frequency mirrors are deployed along with the multiple transmit antennas. Accordingly, least squares and linear minimum mean square error-based channel estimation schemes are proposed for MBM-based systems for the first time in the literature where former studies mainly assume either perfect channel state information or an error model on channel coefficients. In addition, corresponding symbol detection performance is studied. To measure the efficiency of the proposed channel estimation approaches, the theoretical upper bounds on average bit error rate are derived and shown to be well overlapped with the computer simulations for the medium and high signal-to-noise ratio regions. This study is important due to the implementation of channel estimation as well as the theoretical derivation of detection bounds for MBM-based communication systems. 	
\end{abstract}



\section{Introduction}

\IEEEPARstart{S}{patial} modulation (SM) inherently incorporates one extra transmit unit through allocating information bits to the corresponding transmit antenna selection for communication in addition to the transmission of conventionally modulated data symbols \cite{Wen19, Mesleh08}. Thus, the overall spectral efficiency, which is one of the crucial design issue of next generation communication systems, can be increased.

Media-based modulation (MBM), proposed recently \cite{Khandani13}, is one of the novel channel modulation techniques where the wireless channel environment can be manipulated through deployed electronic radio frequency (RF) mirrors on the transmit antennas \cite{Basar19}. Based on the digital on/off condition of controllable RF mirrors, which is specified by the RF source bits, different channel environments are realized. This helps to diversify the channel fading realizations that leads to improved system performance due to converting Rayleigh fading channel to additive white Gaussian noise (AWGN) channel \cite{Mao20}.   

On the other hand, spatial MBM (SMBM), which is the combination of SM and MBM principles belonging to index modulation (IM) family \cite{Ersin19, Basar16}, is able to provide both increased spectral efficiency especially when applied to orthogonal frequency division multiplexing (OFDM)-based systems and channel robustness. Accordingly, there are many studies concentrating on IM family schemes from various perspectives on the system performance \cite{Mao20}-[15]. In \cite{Basar12}, the performance of SM is investigated in the presence of channel estimation errors. In \cite{Gong18}, block pilot-based channel estimation is proposed for generalized SM (GSM)-OFDM systems where the IM idea is extended to frequency domain. Similarly, only detection performance of generalized multiple-mode OFDM with IM is demonstrated in \cite{Wen18}. In \cite{Yildirim17}, quadrature channel modulation is proposed as a combination of quadrature SM and MBM, and, its corresponding detection performance analysis is evaluated. In \cite{Naresh17}, GSM-MBM system detection performance is examined in terms of bit-error-rate (BER). In \cite{Naresh18}, the effect of imperfect channel estimation is studied on the BER performance of MBM by simply introducing an error model on the perfect channel state information (P-CSI). In \cite{Acar19}, least squares (LS)-based channel estimation is applied to IM based OFDM system by employing pilot symbols into the OFDM data. On the other hand, a compressed sensing-based joint user and symbol detection approach is proposed for MBM-enable massive machine-type communications \cite{Yuan20}.

In the studies addressed in [1]-\cite{Yuan20}, the main concentration is to obtain detection performance without performing channel estimation except \cite{Acar19} and \cite{Yuan20} where authors implement channel estimation for OFDM-IM and massive machine-type systems separately, respectively. The common view is to use channel estimation errors to model the performance under imperfect CSI (I-CSI) to get rid of challenging channel estimation problem. 

In this letter, LS and linear minimum mean square error (LMMSE)-based channel estimation schemes are considered for SMBM systems operated under block-wise time-varying Rayleigh fading channels where the effects of channel estimation to the performance of SMBM systems has not been investigated yet. The effect of number of RF mirrors deployed in SMBM on the system performance is investigated. In addition, symbol detection performance is evaluated in terms of BER for different modulation types. Moreover, an upper bound on average BER (ABER) is derived theoretically and shown to be attainable for the proposed channel estimation scheme even in I-CSI conditions. Computer simulation results confirm that channel estimation performance improves with the increasing number of RF mirrors due to the realization of additional channel environments. In the medium-to-high signal-to-noise ratio (SNR) regime, LS-based estimation performance is capable of achieving the LMMSE performance.


\section{
System Model}
\label{sec:guidelines}

\begin{figure*}[h]
	\centering
	\includegraphics[keepaspectratio, width=1\textwidth]{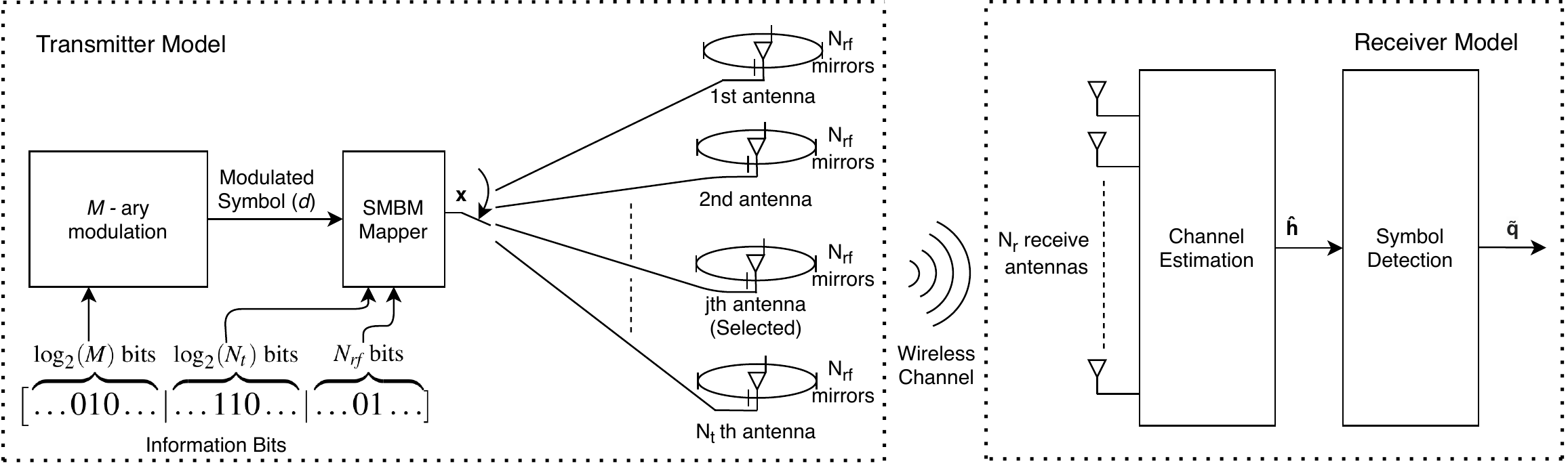}
	\caption{SMBM system model.}
	\label{Fig_SystemModel}	
\end{figure*}

In this section, SMBM system model is introduced by providing transmitter and receiver characteristics. The transceiver model of SMBM system is illustrated in Fig.~\ref{Fig_SystemModel} 
where SMBM system is assumed to have $N_t$ transmit and $N_r$ receive antennas. In addition, each transmit antenna has a number of deployed $N_{rf}$ RF mirrors. Each RF mirror either reflects the incident wave or  passes the incident wave which is called on/off status of RF mirrors. On/off status of an RF mirror can be controlled by one information bit and each on/off combination of RF mirrors creates different radiation patterns.

\subsection{Transmitter Model}
In one communication interval of SMBM systems
\begin{align}
\eta & = \log_2M+\log_2N_t+N_{rf}
\label{Eta}
\end{align}
bits per channel use (bpcu) are generated by the information source and these bits can be classified as 
\begin{eqnarray}
{\bf q}=\big[ \overbrace{\dots 010 \dots}^{\mathclap{\log_2M\text{ bits} }} | \overbrace{\dots 110 \dots }^{\mathclap{\log_2N_t bits}} |  \overbrace{\dots 01 \dots}^{\mathclap{N_{rf}\text{ bits}}} 
],
\label{SourceBits}
\end{eqnarray}
where the first $\log_2M$ bits are used to obtain modulated data signals depending on its modulation type ($M$-quadrature amplitude modulation, $M$-QAM or $M$-phase shift keying, $M$-PSK signal), while $\log_2N_t$ and $N_{rf}$ bits are used to determine the active transmit antenna for SM and the on/off status of RF mirrors (i.e., channel state) for MBM, respectively.

Accordingly, in the mapper block of SMBM transmitter, one transmit antenna and one channel state is selected for transmission by the source bits allocated to determine the active transmit antenna and channel realization, respectively. Since there are possible $N_t$ transmit antennas and $2^{N_{rf}}$ channel states, the transmitted signal vector is represented as
\begin{eqnarray}
	{\bf x}=\big[ \underbrace{\overbrace{0 \dots 0}^{\mathclap{\text{{\it N\textsubscript{t}} zeros}}} | \overbrace{0 \dots 0}^{\mathclap{\text{{\it N\textsubscript{t}} zeros}}} | \dots | \overbrace{0 \dots d \dots 0}^{\mathclap{\text{{\it N\textsubscript{t}}\;-\;1 zeros}}} | \dots | \overbrace{0 \dots 0}^{\mathclap{\text{{\it N\textsubscript{t}} zeros}}}}_{\mathclap{\text{ 2\textsuperscript{{\it N\textsubscript{rf}}} blocks}}} 
	]^T \in \mathbb{C}^{N_t2^{N_{rf}}},
	\label{TransmitVec}
\end{eqnarray}
where the modulated data symbol, $d$ is conveyed at the $m$th coordinate that can be formulated as $m=(k-1)N_t+j$ with $k \in \{1,2,\ldots,2^{N_{rf}}\}$ and $j\in\{1,2,\ldots,N_t\}$ denoting the indices of channel state and active transmit antenna, respectively. Here, $[\cdot]^T$ stands for the transpose operation. Note that transmit signal vector ${\bf x}$ is naturally sparse vector since it consists of one nonzero modulated signal value and $N_t2^{N_{rf}}\!-\!1$ zeros thanks to the SMBM technique.

\subsection{Receiver Model}
The received signal model of SMBM communication system can be represented as
\begin{eqnarray}
{\bf y}={\bf G} {\bf x} + {\bf w},
\label{ReceivedSignal}
\end{eqnarray}
where ${\bf G}$ is the channel matrix comprising of Rayleigh fading coefficients and ${\bf w}$ represents the zero-mean AWGN with variance $\sigma^2_{w}$. The channel matrix can be decomposed as
\begin{eqnarray}
	{\bf G}=\Big[{\bf G}^1, {\bf G}^2, \ldots, {\bf G}^{2^{N_{rf}}}\Big],
	\label{ChanMtxDecomposed}
\end{eqnarray}
where 
\begin{gather}
\label{Sub-channelMtx}
{\bf G}^k\!
=\!
\begin{bmatrix}
\!h_{1,1}^k&\!h_{1,2}^k&\!\dots&\!h_{1,N_t}^k\\
\\\!h_{2,1}^k&\!h_{2,2}^k&\!\dots&\!h_{2,N_t}^k\\
\!\vdots&\!\vdots&\!\ddots&\!\vdots\\
\!h_{N_r,1}^k&\!h_{N_r,2}^k&\!\dots&\!h_{N_r,N_t}^k\\
\end{bmatrix}\!
\end{gather}
is the sub-channel matrix belonging to the $k$th channel state with $h^k_{i,j}$ denoting the channel coefficient between $i$th receiver and $j$th transmitter antennas at the $k$th state for $i\in\{1,2,\ldots,N_r\}$.
 
\section{Channel Estimation and Symbol Detection}
In this section, the channel estimation and symbol detection are explained in SMBM systems. 

We consider block-wise time-varying Rayleigh fading channel where coefficients remain unchanged for a certain communication interval. In order to estimate the channel, a reference signal (i.e., pilot symbol) $p$ with unit energy is used first regarding the all possible channel coefficients. Then, communication still continues for the remaining signaling interval until channel changes to detect the data symbols by using the estimated channel coefficients.

\subsection{Channel Estimation}
The received signal expression defined in \eqref{ReceivedSignal} can be rearranged alternatively in accordance with the reference signal-aided channel estimation process as
\begin{eqnarray}
{\bf r}={\bf P} {\bf h} + {\bf n},
\label{ReceivedSigAlternative}
\end{eqnarray} 
where ${\bf r, P, h}$ and ${\bf n}$ denote the received reference signal vector, transmission matrix, channel vector and zero-mean AWGN with variance 
$\sigma_{n}^2$, respectively. Since same reference signal is used throughout the estimation process and only one symbol is transmitted actively in a signaling duration, reference signal transmission matrix ${\bf P}$ appears as a diagonal matrix. Therefore, transmission matrix can be written as
\begin{eqnarray}
	\label{PilotObsVec}
	{\bf P}
	=
	\left[\begin{array}{ccccc}
		{\it p}&0&\dots&\dots&0\\
		0&{\it p}&0&\dots&0\\
		\vdots&0&\ddots& &\vdots\\
		\vdots&\vdots & &\ddots&0\\
		0&0&\dots&0&{\it p}\\
	\end{array}\right],
\end{eqnarray}
where ${\bf n}$ is the corresponding zero-mean AWGN with variance $\sigma_{n}^2$.
However, channel vector ${\bf h}$ contains all possible channel coefficients and can be expressed as
\begin{eqnarray}
	\label{ChannelVector}
	{\bf h}=\Big[h_{1,1}^1, \ldots,h_{1,N_t}^{2^{N_{rf}}}, h_{2,1}^1\ldots, h_{2,N_t}^{2^{N_{rf}}}, \ldots, h_{N_r,1}^{1}, \ldots, h_{N_r,N_t}^{2^{N_{rf}}}\Big]^T.
\end{eqnarray}

After reference signal transmission process, channel coefficients can be estimated by applying LS and LMMSE considering the linear model defined in \eqref{ReceivedSigAlternative}, respectively, as
\begin{align}
{\bf \hat{h}}&=({\bf P}^\dag {\bf P})^{-1} {\bf P}^\dag {\bf r},
\label{h hat_LS}\\
{\bf \hat{h}}&=({\bf P}^\dag {\bf P}+ {\sigma}_{n}^2 {\bf R}_h^{-1})^{-1} {\bf P}^\dag {\bf r},
\label{h hat}
\end{align}
where ${\bf R}_h=\mathbb{E}[{\bf hh}^\dagger]$ represents the autocorrelation matrix of Rayleigh distributed channel coefficients which is a diagonal matrix whose diagonal components are all equal to channel power, $\sigma_{h}^2$ for this case while ${\bf \hat{h}}$ contains estimated channel coefficients and $[\cdot]^\dagger$ stands for the Hermitian conjugate operation. Note that the diagonal autocorrelation matrix becomes unitary matrix since the channel power is normalized to unity.

When the channel estimation phase is completed, the channel estimation performance can be evaluated in terms of mean square error (MSE) that can be expressed as  
\begin{align}
\textrm{MSE} = {\frac{1}{2^{N_{rf}}\;N_t\;N_r}} \mathbb{E}[({\bf \hat{h}}-{\bf h})^\dag ({\bf \hat{h}}-{\bf h})],
\end{align}
where $\mathbb{E}$ denotes the expectation operation.

\subsection{Symbol Detection}
Symbol detection phase is operated following the channel estimation phase. To make the detection process of transmitted symbols convenient, the channel coefficients between $j$th transmitter and all receivers at $k$th channel state can be expressed as
\begin{eqnarray}
{\bf h}_j^k=[h_{1,j}^k, h_{2,j}^k, \dots, h_{N_r,j}^k]^T.
\label{hjk}
\end{eqnarray}

Accordingly, received signal demodulation is applied using maximum-likelihood (ML) detection. In order to clarify detection phase, let us first define ${\bf {\hat h}}_j^k$ as the estimation of the channel vector, ${\bf h}_j^k$ defined in \eqref{hjk}. Through ML detection, data symbol, transmitter antenna index and RF mirror state are estimated by finding minimum Euclidean distance as represented below
%
\begin{eqnarray}
[\it \tilde{d},\hat{j},\hat{k}] = \arg \min_{\it d,j,k}\Big(||{\bf y}-{\it d}{\bf \hat{h}}_j^k||^2\Big).
\label{EQN_TransmitSignal}
\end{eqnarray}

\section{Average Bit Error Probability Analysis }

In this section, average bit error probability (ABEP), i.e., an upper bound on ABER, for SMBM system is calculated by the derivation of average pair-wise error probability (PEP) for a single receiver case and generalization of this derivation to the systems with multiple receivers. 

\subsection{ABEP for Single Receive Antenna}
In a single receive antenna (SRA) system, probability of error can be defined and analytically calculated as \cite{Mesleh15}
\begin{align}
\nonumber
P_e & = P_r((h_l,s_{\ell})\rightarrow(h_{\tilde{l}},s_{\tilde{\ell}})|{\bf \hat{h}})
\label{prError}
\nonumber\\
~ & = Q \left(\sqrt{\frac{||h_l s_{\ell}-h_{\tilde{l}} s_{\tilde{\ell}}||^2}{2(\sigma_e^2 |s_{\ell}|^2+\sigma_n^2)}}\,\,\right)
= Q\left(\sqrt{{\gamma}}\,\,\right),
\end{align} 
where ${\it h_l}$ represents the $l$th element of channel vector corresponding to the related transmitter antenna and RF mirror combination
while ${\it s_{\ell}}$ and $\sigma_e^2$ denote the ${\it \ell}$th modulated symbol and the variance of disruptive effects in channel estimation, respectively.
Probability density function of ${\gamma}$ can be found as $p_{{\gamma}}({\gamma})={\frac{1}{{\bar{\gamma}}}}\exp({\frac{-{\gamma}}{{\bar{\gamma}}}})$. Using partial derivation method, average PEP for an SRA can be calculated as
\begin{align}
\nonumber
\textrm{PEP}_{\textrm{SRA}}& = \int_{0}^{\infty} P_e\; p_{{\gamma}}(\rho)\; d{\rho}\\
\nonumber
& = \int_{0}^{\infty} Q \left(\sqrt{\frac{||h_l s_{\ell}-h_{\tilde{l}} s_{\tilde{\ell}}||^2}{2(\sigma_e^2 |s_{\ell}|^2+\sigma_n^2)}}\,\,\right) p_{{\gamma}}(\rho)  d{\rho}\\
& = {\frac{1}{2}} \left(1-\sqrt{{\frac{{\bar{\gamma}}/2} {1+{\bar{\gamma}}/2}} }\,\, \right),
\label{PError}
\end{align}
where
\begin{align}
{\bar{\gamma}}& = {\frac{E_s{\sigma}^2_{h}(1+{\sigma}^2_{e})}{2( {\sigma}^2_n + {\sigma}^2_{e}|s_{\ell}|^2)} } \times
\begin{cases}
|s_{\ell} - s_{\tilde{\ell}}|^2,& \text{if } \tilde{l}=l\\
|s_{\tilde{\ell}}|^2 + |s_{\ell}|^2,              & \text{if } \tilde{l} \neq l
\end{cases}.
\end{align} 
Thus, ABER for SRA case in SMBM systems is upper bounded by the ABEP as
\begin{align}
	\label{EQN_ABEP_SRA}
	\textrm{ABER} \leq {\frac{1}{{\eta}2^{\eta}}} \sum_{l,\ell}^{} \sum_{\tilde{l},\tilde{\ell}}^{}
	\textrm{PEP}_{\textrm{SRA}}\;
	e(l,\ell\rightarrow \tilde{l},\tilde{\ell}), 
\end{align}
where $e(l,\ell\rightarrow \tilde{l},\tilde{\ell})$ is number of erroneously detected bits related with $(l,\ell\rightarrow \tilde{l},\tilde{\ell})$ event and the right-hand side of the inequality in \eqref{EQN_ABEP_SRA} corresponds to the closed-form expression of theoretical ABEP for SRA case.

\subsection{ABEP for Multiple Receive Antennas}
In multiple receive antenna (MRA) case, the square of Euclidean distance between two transmitted signals can be written as a sum of squared Euclidean distances between these two transmitted signals in each antenna. Therefore, error probability for a given channel matrix, considering an MRA case can be analytically calculated as
\begin{align}
\nonumber
P(e|{\rho}) & = P_r(({\bf h}_l,s_{\ell})\rightarrow({\bf h}_{\tilde{l}},s_{\tilde{\ell}})|{\bf \hat{G}})\\
\nonumber
~ & = Q \left(\sqrt{\frac{\sum_{i=1}^{N_r}||h_{l,i} s_{\ell}-h_{\tilde{l},i} s_{\tilde{\ell}}||^2}{2(\sigma_e^2 |s_{\ell}|^2+\sigma_n^2)}}\,\,\right)\\
~ & \triangleq Q\left(\sqrt{{\gamma}_{MRA}}\,\,\right),
\label{prErrorMultipleAntenna}
\end{align}
where ${\bf \hat{G}}$ stands for the estimated channel matrix.

The series sum of $\gamma$ variable for an SRA provides a random variable with chi-square distribution for the MRA case. When the chi-square distributed random variable is integrated for all possible values in a similar manner, average PEP for multiple receive antenna can be derived as
\begin{align}
\textrm{PEP}_{\textrm{MRA}}& = \big(\textrm{PEP}_{\textrm{SRA}}\big)^{N_{r}} \sum_{i=0}^{N_{r}-1} {N_r -1 +i\choose i} (1-\textrm{PEP}_{\textrm{SRA}})^i.
\end{align}

Finally, ABER for MRA case in SMBM systems is upper bounded by the ABEP as
\begin{align}
	\label{EQN_ABEP_MRA}
\textrm{ABER} \leq {\frac{1}{{\eta}2^{\eta}}} \sum_{l,\ell}^{} \sum_{\tilde{l},\tilde{\ell}}^{}
\textrm{PEP}_{\textrm{MRA}}\;
e(l,\ell\rightarrow \tilde{l},\tilde{\ell}),
\end{align}
where the right-hand side of the inequality in \eqref{EQN_ABEP_MRA} corresponds to the closed-form expression of theoretical ABEP for MRA case.

\section{Simulation Results}
In the following, Monte-Carlo channel estimation and symbol detection simulation results of SMBM system will be given along with the theoretical ABER for P-CSI and I-CSI, utilizing block-wise time varying uncorrelated Rayleigh fading channel with respect to signal-to-noise ratio (SNR) defined as SNR = $E_b / {\sigma}_{n}^2$, where $E_b$ denotes average energy per bit. Accordingly, channel estimation and symbol detection performances of SMBM systems are provided by taking into account different channel and modulation circumstances for different SNR values.

\begin{figure}[t]
	\centering
	{\includegraphics[width=0.485\textwidth, height=0.5\textheight, keepaspectratio=true]{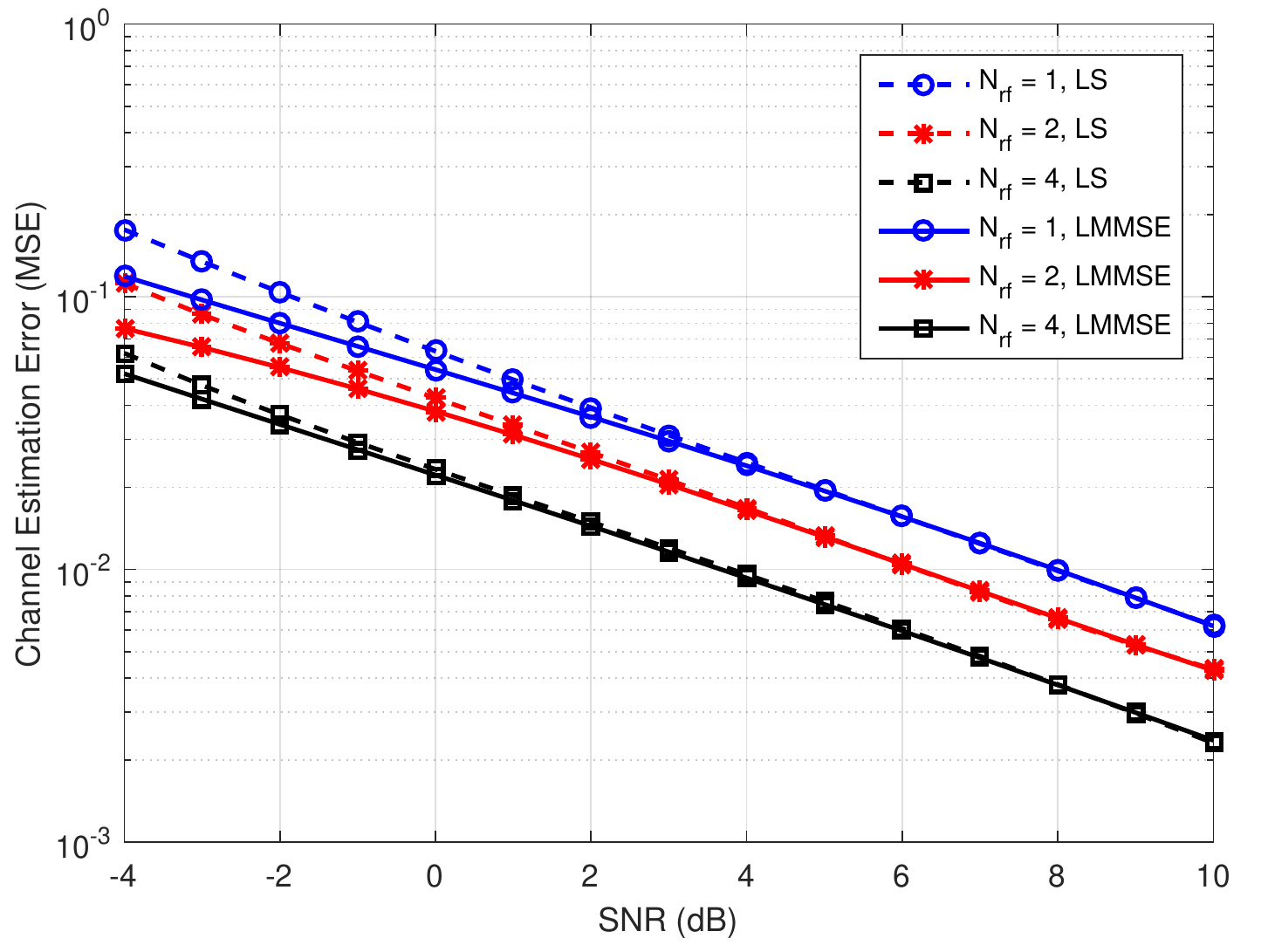}}
	\caption{LS and LMMSE based channel estimation performance of SMBM systems with respect to varying number of RF mirrors.}
	\label{MSE}
\end{figure}

Starting with Fig. \ref{MSE}, 
channel estimation error performances of LS and LMMSE estimation techniques are shown for SMBM system with four transmit antennas, four receive antennas (i.e., $N_t = N_r = 4$) and various numbers of RF mirrors (i.e., $N_{rf} =\{1,2,4\}$). Thanks to the utilization of channel statistics such as correlation and noise variance, LMMSE is superior to LS for the low SNR region (i.e., betwen -4dB and 4dB) where the noise variance is significant. In addition, it can be pointed out that LS and LMMSE based estimation schemes perform better for the systems with higher numbers of RF mirrors due to increase in diversity against fading channel on the transmitter side.

\begin{figure}[t]
	\centering
	{\includegraphics[width=0.485\textwidth, height=0.5\textheight, keepaspectratio=true]{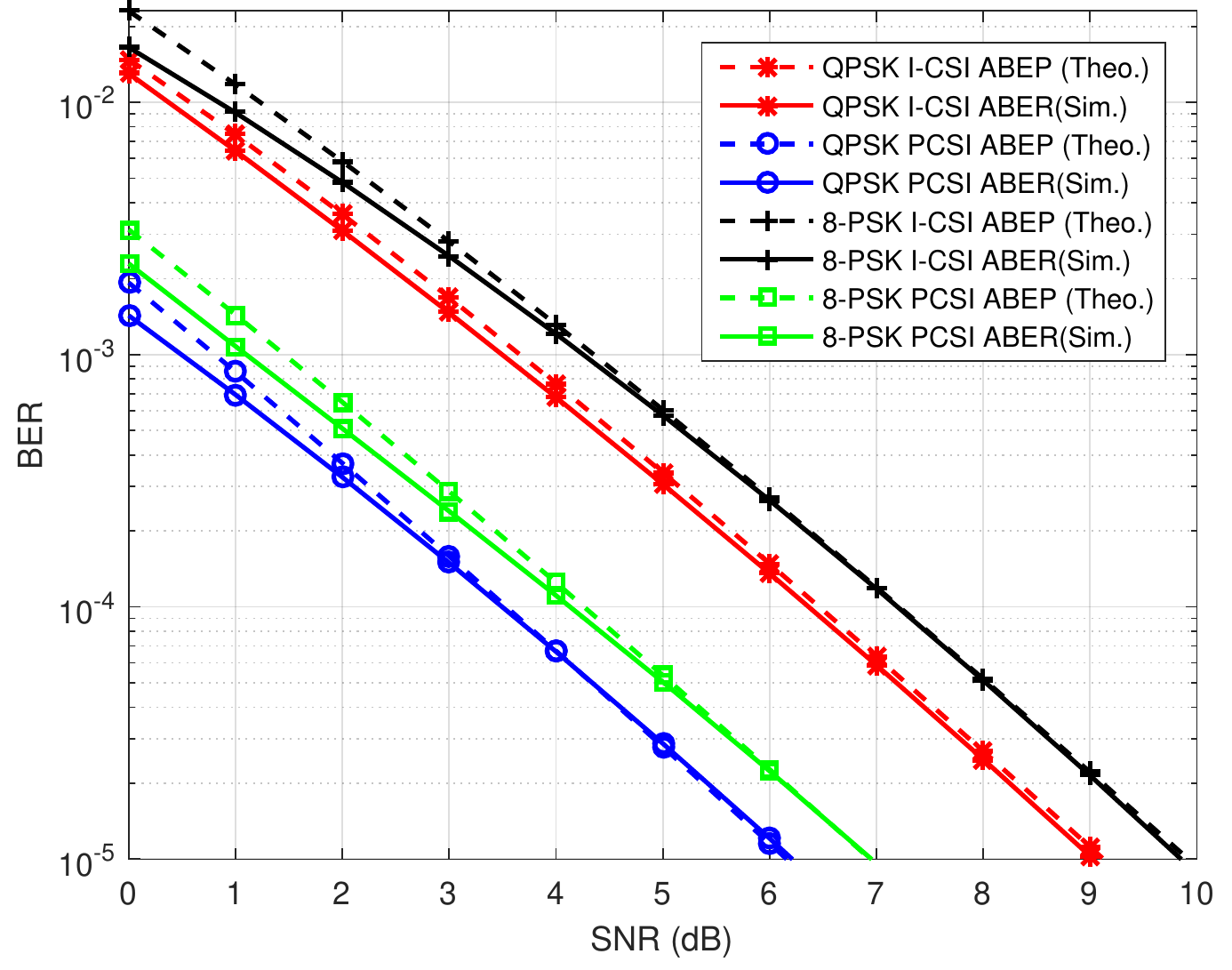}}
	\caption{BER performance of SMBM systems with QPSK and 8-PSK modulation under P-CSI and I-CSI.}
	\label{QPSK and 8-PSK}
\end{figure}

Secondly, symbol detection performance results and analytically calculated ABEP curves using \eqref{EQN_ABEP_MRA} belonging to LMMSE estimation are given for P-CSI conditions (i.e., ideal case) in addition to the I-CSI conditions (i.e., realistic case) in Fig. \ref{QPSK and 8-PSK} 
considering an SMBM system for a quadrature PSK (QPSK) and 8-PSK modulated communication scenario specialized as $N_t = N_r = 4,N_{rf} =2$. In this context, the spectral efficiencies of data transmission utilizing QPSK and 8-PSK modulations are ${\eta} = \log_2(M)+N_{rf}+\log_2(N_t)=6$ bpcu and ${\eta}=7$ bpcu, respectively. Increasing modulation level leads to the detection performance degradation, as expected. In the realistic case (i.e., I-CSI, where channel estimation is applied), detection performance is worse almost 3dB SNR compared to ideal case (i.e., P-CSI) for the same BER value. On the other hand, derived theoretical upper bounds on ABER perfectly match with the simulations for CSI conditions as well as modulation types for a certain region of SNR values.

\begin{figure}[t!]
	\centering
	{\includegraphics[width=0.485\textwidth, height=0.5\textheight, keepaspectratio=true]{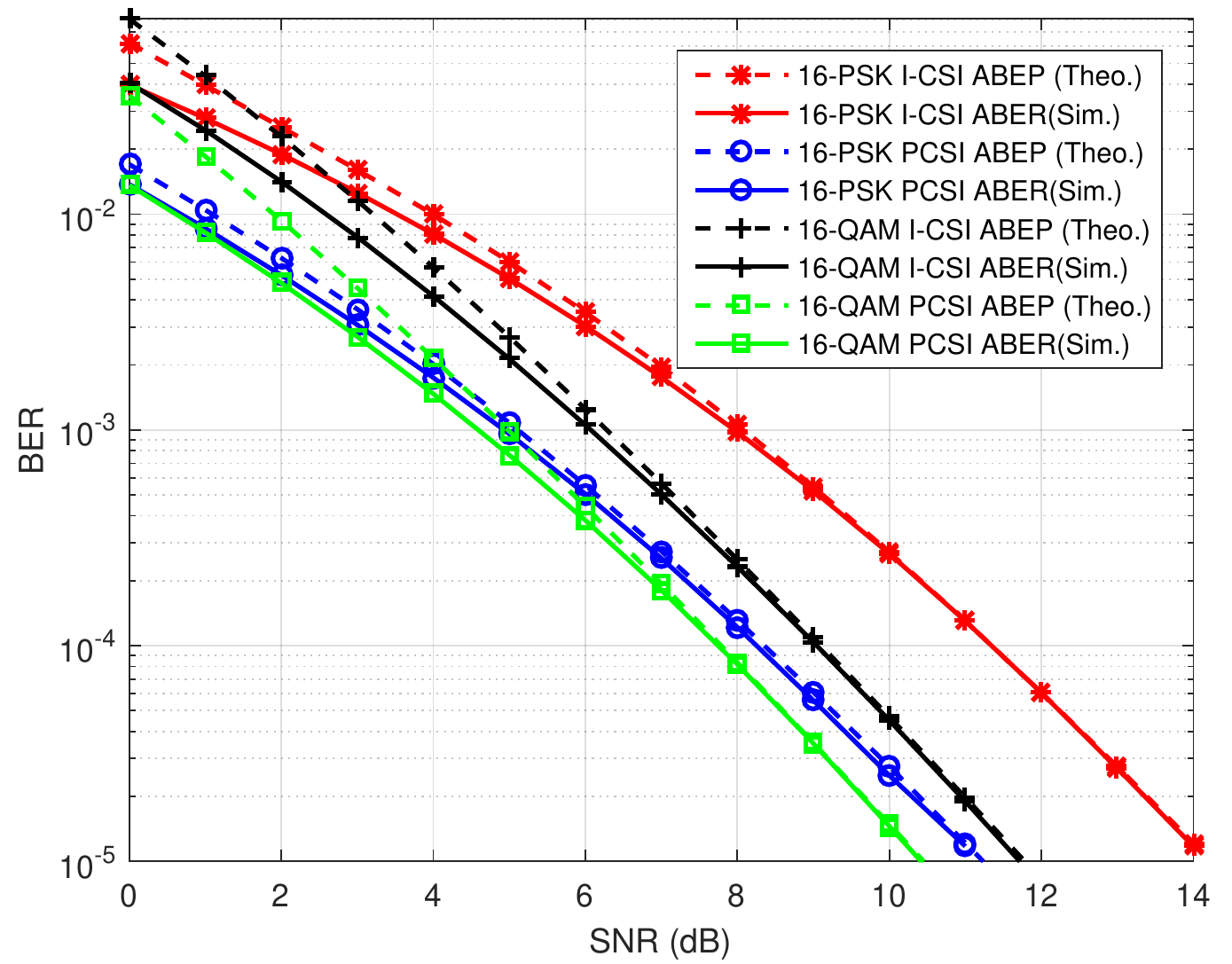}}
	\caption{BER performance of SMBM systems with 16-PSK and 16-QAM modulation under P-CSI and I-CSI.}
	\label{16-PSK and 16-QAM}
\end{figure}

Third, BER performances for 16-level modulations (16-QAM and 16-PSK) resulting in 8 bpcu utilization are shown in Fig. \ref{16-PSK and 16-QAM}. 
Similarly, theoretical ABEP curves are tight bounds for ABER curves. In addition, BER results of 16-QAM are better than 16-PSK due to the higher distance among constellation points. Furthermore, compared to 8-PSK, it is also worth noting that, there is an around 2.5dB performance gap on BER values of 16-PSK among P-CSI and I-CSI cases.


\section{Conclusion}
In this letter, LS and LMMSE based channel estimation schemes for SMBM systems under a Rayleigh fading channel are first proposed to demonstrate SMBM communication system performance in a more realistic manner. In addition, a theoretical upper bound for ABER is derived and shown to be the tight bound for Monte-Carlo simulation results. 
The effects of number of RF mirrors deployed on the transmit antennas and modulation types are investigated in detail. An increase in RF mirror deployment provides better channel estimation results thanks to the diversity in MBM. Moreover, detection performance of I-CSI case is found to be worse than P-CSI case with a gap of 2-3dB.
These outcomes are important for providing a practical channel estimation perspective to SMBM communication system performance.

\ifCLASSOPTIONcaptionsoff
\newpage
\fi

\bibliographystyle{IEEEtran}
\bibliography{Bibliography_IEEE_TVT_AMH}

\begin{thebibliography}{10}
\providecommand{\url}[1]{#1}
\csname url@samestyle\endcsname
\providecommand{\newblock}{\relax}
\providecommand{\bibinfo}[2]{#2}
\providecommand{\BIBentrySTDinterwordspacing}{\spaceskip=0pt\relax}
\providecommand{\BIBentryALTinterwordstretchfactor}{4}
\providecommand{\BIBentryALTinterwordspacing}{\spaceskip=\fontdimen2\font plus
\BIBentryALTinterwordstretchfactor\fontdimen3\font minus
  \fontdimen4\font\relax}
\providecommand{\BIBforeignlanguage}[2]{{%
\expandafter\ifx\csname l@#1\endcsname\relax
\typeout{** WARNING: IEEEtran.bst: No hyphenation pattern has been}%
\typeout{** loaded for the language `#1'. Using the pattern for}%
\typeout{** the default language instead.}%
\else
\language=\csname l@#1\endcsname
\fi
#2}}
\providecommand{\BIBdecl}{\relax}
\BIBdecl

\bibitem{Wen19}
M.~{Wen}, B.~{Zheng}, K.~J. {Kim}, M.~{Di Renzo}, T.~A. {Tsiftsis}, K.~{Chen},
  and N.~{Al-Dhahir}, ``A survey on spatial modulation in emerging wireless
  systems: Research progresses and applications,'' \emph{IEEE J. Sel. Areas
  Commun.}, vol.~37, no.~9, pp. 1949--1972, 2019.

\bibitem{Mesleh08}
R.~Y. {Mesleh}, H.~{Haas}, S.~{Sinanovic}, C.~W. {Ahn}, and S.~{Yun}, ``Spatial
  modulation,'' \emph{IEEE Trans. Veh. Technol.}, vol.~57, no.~4, pp.
  2228--2241, Jul. 2008.

\bibitem{Khandani13}
A.~K. {Khandani}, ``Media-based modulation: A new approach to wireless
  transmission,'' in \emph{IEEE Int. Symp. Inf. Theory}, Jul. 2013, pp.
  3050--3054.

\bibitem{Basar19}
E.~{Basar}, ``Media-based modulation for future wireless systems: A tutorial,''
  \emph{IEEE Wireless Commun.}, vol.~26, no.~5, pp. 160--166, 2019.

\bibitem{Mao20}
T.~{Mao}, Q.~{Wang}, M.~{Wen}, and Z.~{Wang}, ``Secure
  single-input-multiple-output media-based modulation,'' \emph{IEEE Trans. Veh.
  Technol.}, vol.~69, no.~4, pp. 4105--4117, 2020.

\bibitem{Ersin19}
E.~Ozturk, E.~Basar, and H.~A. Cirpan, ``Multiple-input multiple-output
  generalized frequency division multiplexing with index modulation,''
  \emph{Phy. Commun.}, vol.~34, pp. 27 -- 37, 2019.

\bibitem{Basar16}
E.~{Basar}, ``{Index modulation techniques for 5G wireless networks},''
  \emph{IEEE Commun. Mag.}, vol.~54, no.~7, pp. 168--175, Jul. 2016.

\bibitem{Basar12}
E.~{Basar}, U.~{Aygolu}, E.~{Panayirci}, and H.~V. {Poor}, ``Performance of
  spatial modulation in the presence of channel estimation errors,'' \emph{IEEE
  Commun. Lett.}, vol.~16, no.~2, pp. 176--179, 2012.

\bibitem{Gong18}
B.~{Gong}, L.~{Gui}, S.~{Luo}, Y.~L. {Guan}, Z.~{Liu}, and P.~{Fan}, ``Block
  pilot based channel estimation and high- accuracy signal detection for
  {GSM-OFDM} systems on high-speed railways,'' \emph{IEEE Trans. Veh.
  Technol.}, vol.~67, no.~12, pp. 11\,525--11\,536, 2018.

\bibitem{Wen18}
M.~{Wen}, Q.~{Li}, E.~{Basar}, and W.~{Zhang}, ``Generalized multiple-mode
  {OFDM} with index modulation,'' \emph{IEEE Trans. Wireless Commun.}, vol.~17,
  no.~10, pp. 6531--6543, 2018.

\bibitem{Yildirim17}
I.~{Yildirim}, E.~{Basar}, and I.~{Altunbas}, ``Quadrature channel
  modulation,'' \emph{IEEE Wireless Commun. Lett.}, vol.~6, no.~6, pp.
  790--793, Dec. 2017.

\bibitem{Naresh17}
Y.~{Naresh} and A.~{Chockalingam}, ``On media-based modulation using rf
  mirrors,'' \emph{IEEE Trans. Veh. Technol.}, vol.~66, no.~6, pp. 4967--4983,
  Jun. 2017.

\bibitem{Naresh18}
------, ``Performance analysis of media-based modulation with imperfect channel
  state information,'' \emph{IEEE Trans. Veh. Technol.}, vol.~67, no.~5, pp.
  4192--4207, May 2018.

\bibitem{Acar19}
Y.~Acar, S.~Aldirmaz-Colak, and E.~Basar, ``{Channel estimation for OFDM-IM
  systems},'' \emph{Turkish J. Elec. Eng. \& Comp. Sci.}, vol.~27, pp.
  1908--1921, 2019.

\bibitem{Yuan20}
X.~{Ma}, S.~{Guo}, and D.~{Yuan}, ``Improved compressed sensing-based joint
  user and symbol detection for media-based modulation-enabled massive machine-
  type communications,'' \emph{IEEE Access}, vol.~8, pp. 70\,058--70\,070,
  2020.

\bibitem{Mesleh15}
R.~{Mesleh}, S.~{Ikki}, and H.~M. {Aggoune}, ``Quadrature spatial modulation,''
  \emph{IEEE Trans. Veh. Technol.}, vol.~64, no.~6, pp. 2738--2742, Jun. 2015.

\end{thebibliography}

\end{document}